\documentstyle[12pt,a4wide]{article}
\normalsize
\def\simless{\mathbin{\lower 3pt\hbox{$\rlap{\raise 5pt\hbox{$\char'074$}}
\mathchar"7218$}}}
\def\simgreat{\mathbin{\lower 3pt\hbox{$\rlap{\raise 5pt \hbox{$\char'076$}}
\mathchar"7218$}}}
\catcode`@=11

\def\beqra{\begin{eqnarray}} \def\eeqra{\end{eqnarray}}
\def\beq{\begin{equation}}      \def\eeq{\end{equation}}

\def\fo{\hbox{{1}\kern-.25em\hbox{l}}}

\def\ch{\@startsection{section}{1}{\z@}{-3ex plus-1ex minus-.2ex}%
        {2ex plus.2ex}{\large\sc}}

\def\; \lapp \;{\raisebox{-.4ex}{\rlap{$\sim$}} \raisebox{.4ex}{$<$}}

\def\con{\ifmmode \hbox{\bf*} \else{\bf*}\fi}   % conjugation
\def\scon{\ifmmode \hbox{\footnotesize\rm\bf*} \else{\footnotesize\rm\bf*}\fi}

\def\0#1{\relax\ifmmode\mathaccent"7017{#1}%    % puts a little circle atop,
        \else\accent23#1\relax\fi}              % as a halo of a saint

\def\eslash{\not{\hbox{\kern-2pt $E$}}}

\catcode`@=12

\begin{document}
\hoffset=0.4cm
\voffset=-1truecm
\normalsize

%%%%%%%%%%%%%%%%%%%%%%%%%%%%

\begin{titlepage}
\begin{flushright}
DFPD 94/TH/38\\
SISSA 94/81-A\\
\end{flushright}
\vspace{12pt}
\centerline{\Large {\bf On the Spontaneous CP Breaking at Finite
Temperature}}
\vskip 0.1 cm
\centerline{\Large {\bf in a Nonminimal Supersymmetric Model}}
\vspace{10pt}
\begin{center}
{\large\bf D. Comelli$^{a,b,}$\footnote{Supported by the
Schweizerischen National Funds and on leave of absence from
Dipartimento di Fisica Teorica, Universit\`a di Trieste, Italy;
Email: comelli@mvxtst.ts.infn.it}
, M. Pietroni$^{c,}$\footnote{Address after September 1, 1994:
DESY, Deutsches Elektronen-Synchroton; Email: pietroni@mvxpd5.pd.infn.it}
 and A. Riotto$^{c,d,}$\footnote{On leave of absence from
International School for Advanced Studies (ISAS), Trieste, Italy;
Email:riotto@tsmi19.sissa.it}}
\end{center}
\vskip 0.2 cm
{\footnotesize
\centerline{\it $^{(a)}$ Institut f\"ur Theoretische Physik Universitat,}
\centerline{\it Wintherthurer Strasse 190, Zurich, Switzerland}
\vskip 0.2 cm
\centerline{\it $^{(b)}$ Istituto Nazionale di Fisica Nucleare,}
\centerline{\it Sezione di Trieste, 34014 Trieste, Italy}
\vskip 0.2 cm
\centerline{\it $^{(c)}$Istituto Nazionale di Fisica Nucleare,}
\centerline{\it Sezione di Padova, 35100 Padua, Italy}
\vskip 0.2 cm
\centerline{\it $^{(d)}$Instituto de Estructura de la Materia,CSIC,}
\centerline{\it Serrano, 123, E-20006 Madrid, Spain}}
\baselineskip=20pt
\vskip 0.5 cm
\centerline{\large\bf Abstract}
\vskip 0.2 cm

We study the spontaneous CP breaking at finite temperature in the Higgs
sector in the Minimal Supersymmetric Standard Model with a gauge
singlet. We consider the contribution of the standard model particles
and that of stops, charginos, neutralinos, charged and neutral Higgs
boson to the one-loop effective potential. Plasma effects for all bosons
are also included. Assuming CP conservation at zero temperature,
so that experimental constraints coming from, {\it e.g.},
the electric dipole moment of the neutron are avoided,  and the
electroweak phase transition to be of the first order and proceeding via
bubble nucleation, we show that spontaneous CP breaking cannot occur
inside the bubble mainly due to large effects coming from the Higgs
sector. However, spontaneous CP breaking can be present in the
region of interest for the generation of the baryon asymmetry, namely inside
the
bubble wall. The important presence of very tiny explicit CP violating
phases is also commented.
\end{titlepage}
\baselineskip=24pt
{\Large {\bf 1. Introduction}}

\vskip 0.8 cm

It is well known that for generating the observed baryon asymmetry
of the Universe (BAU) three basic ingredients are necessary \cite{sak}:
baryonic violating interactions, departure from thermal equilibrium and
C and CP violation. Even if these conditions can be esaily fulfilled
in grand unified theories (GUT's) \cite{gut}, severe complications are
present. First of all, any non zero fermion number $(B+L)$ created at
some superheavy scale is almost completely arised by the anomalous
electroweak $(B+L)$ violating processes \cite{kuz}, which are in
equilibrium down to a temperature of about $10^{2}$ GeV. Moreover, if
$(B-L)$ violating processes are present and in equilibrium at high
temperatures, then also an eventual $(B-L)$ component of the BAU
vanishes before the onset of the electroweak era, unless severe
constraints on the $(B-L)$ violating couplings are imposed \cite{L} or
ad hoc ways out are found \cite{way}.

All these considerations make the possibility of generating the BAU
during the electroweak phase transition (EWPT) very appealing
\cite{nelson}. Here the difficulty is threefold. First of all, it is
still an open question if the necessary amount of CP violation to
produce enough baryon asymmetry is present in the standard model (SM)
\cite{if}; secondly, it is not obvious that the phase transition is
enough strongly first order to make the baryon number production
effective; third, one must avoid the wiping out of the baryon asymmetry
by anomalous processes, which leads to an upper bound on the mass of the
SM Higgs boson \cite{bound} already ruled out by LEP results \cite{lep}.

The situation does not appear much more appealing when the minimal
supersymmetric extension of the standard model (MSSM) \cite{haber} is
considered. The MSSM contains two extra CP violating phases with respect
to the SM. The requirement that these phases provide the necessary
amount of CP violation for the generation of the BAU gives rise to
additional strong constraints on the parameter space of the model
\cite{s}. Indeed, the electric dipole moment of the neutron must be
larger than $10^{-27}$ e-cm, while an improvement of the current
experimental bound on it by one order of magnitude would constrain the
lightest chargino and the lightest neutralino to be lighter than 88 and
44 GeV, respectively.
As far as the spontaneous CP breaking (SCPB) in the MSSM is
concerned, it can be triggered by radiative corrections \cite{mae}
at zero temperature. As
established on general grounds by Georgi and Pais \cite{gp}, it requires
the existence of a pseudoscalar Higgs boson with a mass of a few GeV
\cite{pom}, which has been ruled out by LEP \cite{lepp}. Nevertheless,
it has been recently realized \cite{cp} that temperature effects can
trigger SCPB in the MSSM at the critical temperature of the EWPT and
that the right amount of the baryon asymmetry can be generated
with the help of tiny CP violating phases giving rise to a neutron
electron dipole moment well below its present experimental limit
\cite{cpr}. Even if this mechanism can work in a wide region of the
parameter space without any fine tuning, it requires, as a general
tendency, small values of the mass of the pseudoscalar Higgs boson and
large values of $\tan\beta=v_2/v_1$, $v_1$ and $v_2$ being the vacuum
expectation values (VEV's) of the two Higgs doublets present in the
model, whereas recent results on the phase transition in the MSSM
\cite{rec} seem to point out towards the opposite direction in order not
to wash out the generated baryon asymmetry.

In the framework of supersymmetric grand unified theories, the MSSM
is not the most general low energy manifestation of supersymmetric
GUT's. Indeed, it is possible that at low energy the theory contains
and additional gauge singlet field, the so-called next-to-minimal
supersymmetric standard model (NMSSM) \cite{n}, as predicted in many
superstring models based on $E_6$ \cite{e6} and $SU(5)\otimes U(1)$
\cite{str} GUT groups. Incidentally, the presence of an additional gauge
singlet allows to get rid of the so called $\mu$ problem in the MSSM,
namely the presence of an unknown supersymmetric mass term $\mu$
in its superpotential.

The study of the EWPT in the NMSSM has been performed in ref.
\cite{max} where it has been shown that the order of the transition is
determined by the trilinear soft supersymmetric breaking terms and that
it is possible to preserve the baryon asymmetry from the wiping out
of sphaleron interactions for masses of the lightest scalar
well above the experimental limit.

In this papar we want to address the question of the SCPB in the NMSSM
at finite temperature, which is one of the key ingredients to generate
the BAU at the electroweak phase transition. Indeed, in Supergravity
inspired models with canonical Kahler potentials it is expected that
CP violation in the parameters of the Higgs potential is not very significant
\cite{alguila} and the presence of only explicit CP violation is
not enough to produce the BAU,
so that any CP violation in the Higgs sector must be
spontaneous. Moreover, it has been shown by Romao \cite{romao} that the
potential of the NMSSM at the tree level and at zero temperature
has no CP violating minimum. The point is that at the CP violating
extremum there is always a mode with negative mass squared. When one-loop
corrections to the tree level potential at zero temperature are added,
the CP violating saddle point is turned into an absolute minimum
\cite{bb}, where some of the Higgs fields are necessarily very light
since they correspond to the modes having negative mass squared when
one-loop corrections are not present. Quantitatively, the presence of a
CP violating minimum requires two neutral and one charged Higgs bosons
to have masses smaller than $\sim$ 100 GeV, which rules out much of the
parameter space \cite{bb}. The aim of this paper is to investigate
whether finite temperature effects can trigger SCPB in the NMSSM in
regions of the parameter space which at zero temperature correspond to
{\it non} CP violating minima, {\it i.e.} to mass eigenvalues for the
lightest Higgses not dangerously light.

The paper is organized as follow. In Section {\bf 2} we discuss the SCPB
in the Higgs sector of the NMSSM on very general grounds, {\it i.e.}
using the {\it most general} gauge invariant potential containing two
Higgs doublets and the gauge singlet. In the same Section we also show
how to renormalize the parameters present in the model giving all the
details in the Appendix. This general discussion will make easier to
understand why at zero temperature CP cannot be broken spontaneously at
the tree level and why, including zero temperature one-loop corrections,
SCPB can occur only in a very restricted region of the parameter space.
In Section {\bf 3} we introduce the corrections coming from the finite
temperature effective potential. In Section {\bf 4} we
 present and discuss our results
making use of what learned in Section {\bf 2}.
 Finally, Section {\bf 5} contains our conclusions.
{\large {\bf 2. Spontaneous CP Violation in the NMSSM}}

\vskip 0.8 cm
{\bf 2.1. General Analysis}
\vspace{0.8cm}\\

The superpotential involving the superfields $\hat{H}_1$, $\hat{H}_2$ and
$\hat{N}$ in the NMSSM is
\begin{equation}
W=\lambda\hat{H}_1\hat{H}_2\hat{N}-\frac{1}{3}k\hat{N}^3+
h_t \hat{Q}\hat{H}_2\hat{U}^c,
\end{equation}
where the $\hat{N}^3$ term is present to avoid a global $U(1)$ symmetry
corresponding to $\hat{N}\rightarrow\hat{N}{\rm e}^{i\theta}$ and
$\hat{H}_1\hat{H}_2\rightarrow \hat{H}_1\hat{H}_2{\rm e}^{-i\theta}$,
and
$\hat{Q}$ and $\hat{U}^c$ denote respectively the left-handed quark
doublet and the (anti) right handed quark singlet of the third
generation. Note that a $Z_3$ symmetry under which any Higgs superfield
$\hat{\phi}$ transforms as $\hat{\phi}\rightarrow\alpha\hat{\phi}$
with $\alpha^3=1$ is still present.
This is also the typical structure
emerging from superstring inspired scenarios.
This symmetry, when spontaneously broken in the vacuum,
could create a serious cosmological problem due to the appearance of
domain walls. However, it has been shown in ref. \cite{kari} that
nonrenormalizable terms like $\sim N^4/M$, $M$ is some superheavy scale
as the GUT or the Planck scale, would prevent the density
of domain walls from becoming large enough to create any cosmological
danger, while being too small to have any impact in the low energy
phenomenology as well as in the following discussions.

The tree level potential is given by
\begin{eqnarray}
V_{tree}&=&V_F+V_D+V_{soft},\nonumber\\
V_F&=&\left|\lambda\right|^2\left[\left|N\right|^2\left(
\left|H_1\right|^2+\left|H_2\right|^2\right)+\left|H_1
H_2\right|^2\right]+\left|k\right|^2\left|N\right|^4\nonumber\\
&-&\left(\lambda k^{*} H_1 H_2
N^{*2}+\mbox{h.c.}\right),\nonumber\\
V_D&=&\frac{1}{8}\left(g^2+g^{\prime 2}\right)\left(
\left|H_1\right|^2-\left|H_2\right|^2\right)^2+\frac{1}{2}
g^2\left|H_{1}^{\dag}H_2\right|^2,\nonumber\\
V_{soft}&=& m_1^2\left|H_1\right|^2 +
m_2^2\left|H_2\right|^2 +
m_N^2\left|N\right|^2\nonumber\\
&-&\left(\lambda A_{\lambda}H_1 H_2 N + \mbox{h.c.}\right)-
\left(\frac{1}{3} k A_k N^3 +\mbox{h.c.}\right),
\end{eqnarray}
where
\begin{equation}
H_1\equiv\left( \begin{array}{c}
H_1^0\\H^{-}
\end{array}\right),\:\:\:\:\:
H_2\equiv \left( \begin{array}{c}
H^{+}\\H_2^{0}
\end{array}\right)
\end{equation}
and $g$ and $g^{\prime}$ are the gauge couplings of $SU(2)_L$
and $U(1)_Y$, respectively.

Redifining the global phases of $H_2$ and $N$, it can be shown that all
the parameters in eq. (2) can be made real, except the ratio
$r=A_{\lambda}/A_k$. We assume this parameter to be real \cite{bb}.
Indeed, in Supergravity models with a canonical Kahler potential
it turns out that $r=1$ and then real at the
unification scale. $r$ can develop a phase through the renormalization
group equations thanks to complex gaugino masses, but this effect is
small due to the constraints on the gaugino phase coming from the
electric dipole moment\footnote{Note however that the presence of even
very small phases for $r$ is crucial for the generation of the BAU, see
\cite{cpr} and below.}.

If we define
\begin{equation}
\langle H_1^{0}\rangle\equiv v_1\: {\rm e}^{i\theta_1},\:\:\:\:
\langle H_2^{0}\rangle\equiv v_2 \:{\rm e}^{i\theta_2},\:\:\:\:
\langle N\rangle\equiv x \:{\rm e}^{i\theta_3},
\end{equation}
and
\begin{equation}
3\theta_3\equiv 2\varphi_3,\:\:\:\:
\theta_1+\theta_2+\theta_3\equiv2\varphi_1,\:\:\:\:
\theta_1+\theta_2-2\theta_3\equiv2\varphi_1-2\varphi_3,
\end{equation}
we can write the most general gauge invariant (under
$SU(2)_L\otimes U(1)_Y\otimes Z_3$) potential in the vacuum
\begin{eqnarray}
\langle V\rangle&=& D_1 v_1^4+D_2 v_2^4 +D_3 v_1^2 v_2^2+D_4 v_2^2 x^2
+D_5 v_2^2 x^2 + D_6 x^4\nonumber\\
&+&D_7v_1v_2x^2\:{\rm cos}\left(2\varphi_1-2\varphi_3\right)+
D_8 m_1^2 v_1^2+ D_9 m_2^2 v_2^2 +D_{10} m_N^2 x^2\nonumber\\
&+&D_{11}v_1 v_2 x\:{\rm cos}\left(2\varphi_1\right)+
D_{12}x^3\:{\rm cos}\left(2\varphi_3\right),
\end{eqnarray}
where it is easy to derive the $D$ coefficients in the case of the tree
level potential just comparing eqs. (2) and (6).

The potential which depends on the phases
\begin{equation}
V\left(\varphi_1,\varphi_3\right)=D_7v_1v_2x^2\:{\rm
cos}\left(2\varphi_1-2\varphi_3\right)+
D_{11}v_1 v_2 x\:{\rm cos}\left(2\varphi_1\right)+
D_{12}x^3\:{\rm cos}\left(2\varphi_3\right),
\end{equation}
is minimized by the following values of $\varphi_1$ and $\varphi_3$
\cite{romao,bb}
\begin{eqnarray}
{\rm cos}\left(2\varphi_1\right)&=&
\frac{1}{2}\left(\frac{A_{13}A_{23}}{A_{12}^2}-\frac{A_{23}}{A_{13}}
-\frac{A_{13}}{A_{23}}\right),\nonumber\\
{\rm cos}\left(2\varphi_3\right)&=&
\frac{1}{2}\left(\frac{A_{13}A_{12}}{A_{23}^2}-\frac{A_{12}}{A_{13}}
-\frac{A_{13}}{A_{12}}\right),\nonumber\\
{\rm cos}\left(2\varphi_1-2\varphi_3\right)&=&
\frac{1}{2}\left(\frac{A_{12}A_{23}}{A_{13}^2}-\frac{A_{12}}{A_{23}}
-\frac{A_{23}}{A_{12}}\right),
\end{eqnarray}
where
\begin{equation}
A_{12}\equiv D_{11}v_1 v_2 x,\:\:, A_{13}\equiv D_7 v_1 v_2 x^2,\:\:
A_{23}\equiv D_{12} x^3,
\end{equation}
if $\left|A_{12}A_{13}\right|$, $\left|A_{12}A_{23}\right|$ and
$\left|A_{13}A_{23}\right|$ form a triangle and if
\begin{equation}
\frac{A_{12}A_{13}}{A_{23}}>0.
\end{equation}
Defining now the new variable \cite{romao}
\begin{equation}
\Sigma\equiv v_1^2+v_2^2,\:\:\:\: \Delta\equiv v_1^2-v_2^2,
\end{equation}
one can show that the extremum in the full parameter space $\left(
\Sigma,\Delta,\varphi_1,\varphi_3\right)$ is a minimum if the following
3x3 matrix is definite positive
\begin{equation}
B=
\left( \begin{array}{ccc}
2B_4& B_6& B_7\\
B_6& 2B_5& B_8\\
B_7& B_8& 2B_9
\end{array}\right),
\end{equation}
where
\begin{eqnarray}
B_4&=& \frac{D_1+D_2}{4}-\frac{1}{4}D_3+\frac{1}{8}
\frac{D_7 D_{11}}{D_{12}},\nonumber\\
B_5&=& \frac{D_1+D_2}{4}+\frac{1}{4}D_3-\frac{1}{8}
\frac{D_7 D_{11}}{D_{12}},\nonumber\\
B_6&=&\frac{D_1-D_2}{2},\nonumber\\
B_7&=&\frac{D_4-D_5}{2},\nonumber\\
B_8&=&\frac{D_4+D_5}{2},\nonumber\\
B_9&=&D_6-\frac{1}{2}\frac{D_7D_{12}}{D_{11}}.
\end{eqnarray}
For the tree level case, we have the particular values for the $B$
parameters
\begin{eqnarray}
B_4&=&-\frac{1}{4}\lambda^2+\frac{1}{8}\left(g^2+g^{\prime 2}\right)-
\frac{3}{4}\lambda^2 r,\nonumber\\
B_5&=&\left(\frac{1}{4}+\frac{3}{4}r\right)\lambda^2,\nonumber\\
B_6&=&B_7=0,\nonumber\\
B_8&=&\lambda^2,\nonumber\\
B_9&=&k^2\left(1+\frac{1}{3r}\right).
\end{eqnarray}
It is now easy to understand why CP cannot be broken spontaneously at
the tree level. Indeed, one has to satisfy the following system of
conditions
\begin{equation}
\frac{A_{12}A_{13}}{A_{23}}>0\Rightarrow r<0\:\:{\rm plus}\:\:
\left\{ \begin{array}{c}
B_9>0\Rightarrow r<-\frac{1}{3}\: {\rm or}\: r>0,\\
B_5>0 \Rightarrow r>-\frac{1}{3},\\
4B_5 B_9-B_8^2>0\Rightarrow (1+3r)^2/3r>\lambda^2/k^2>0,
\end{array}\right.
\end{equation}
which does not have any solutions in the $r$ space. The key point here
is the following: once the condition given in eq. (10) is satisfied,
which assures
that the nonvanishing phases in eq. (8) correspond to a minimum,
then the $B$ matrix is never definite positive at the tree level. When
one-loop corrections are considered, while the parameters involving
the condition (10) receive small corrections because the latter are
always proportional to $\lambda$ or $k$ which are taken to be small (see
below),
among the elements of $B$, $B_5$ may receive
very large corrections from the top-stop sector and the Higgs
sector (see the discussion of Section {\bf 4}).
It is then clear that, to have SCPB, the correction to the tree level
potential should act predominantly on the $B_5$ term to drive it and
$\left(4B_5 B_9-B_8^2\right)$ to positive values. This fact will be
crucial for the following and expecially for the
discussion about the finite temperature corrections to the tree level
potential.
\vspace{0.8cm}

{\bf 2.2. How to Renormalize the Tree Level Potential}

\vspace{0.8cm}

In this Subsection we want to give some details on how to calculate
the corrections to the tree level coefficients of the most general
potential, eq. (6). A complete full analysis will be given in the
Appendix. Note also that the following discussion can be extended to any
model and to any potential.

The one-loop effective potential can be written as
\begin{equation}
V_{eff}=V_{tree}+V_1,
\end{equation}
where $V_{1}\left(H_1^0,H_2^0,N\right)$ can indicate either the one-loop
correction to the tree level potential at zero temperature or the
one-loop correction at finite temperature. The
$SU(2)_L\otimes U(1)_Y\otimes Z_3$ invariant and independent operators
in the NMSSM are the dimension two operators $\left|H_1\right|^2$,
$\left|H_2\right|^2$ and $\left|N\right|^2$ and the dimension three
operators 2 Re$\left(N^3\right)$, 2 Re$\left(H_1H_2N^{* 2}\right)$ and
2 Re$\left(H_1H_2N\right)$. Let's collectively indicate them
with $\varepsilon_i$ and $\epsilon_i$, respectively.
The effective potential $V_{eff}$ can be then
written as follows
\[
V_{eff}=\sum_i
a_i\varepsilon_i+\sum_{i<j}a_{ij}\varepsilon_{i}\varepsilon_{j}
+\sum_i b_i\epsilon_i,
\]
where $a_i$, $a_{ij}$ and $b_i$ refer to dimension two, four and three
operators, respectively.

Since $V_1$ receives its contributions from all the particles whose
masses are field dependent and therefore is a function of such masses,
one can easily calculate the coefficients of each gauge invariant
operator in the following way (formulas for the dimension two operators
are analogous to those for the dimension three operators)

$\bullet$ dimension three operators:
2 Re$\left(N^3\right)$, 2 Re$\left(H_1H_2N^{* 2}\right)$ and
2 Re$\left(H_1H_2N\right)$
\[
b_i=\left.\frac{\partial V_{eff}}{\partial \epsilon_i}\right|_0=
\left.\frac{\partial V_{tree}}{\partial \epsilon_i}\right|_0
+\left.\sum_a\frac{\partial V_1}{\partial m_a^2}\frac{
\partial m_a^2}{\partial\epsilon_i}\right|_0,
\]

$\bullet$ dimension four operators: $\left|H_1\right|^4$,
$\left|H_2\right|^4$, $\left|N\right|^4$,
$\left|H_1\right|^2\left|H_2\right|^2$,
$\left|H_1\right|^2\left|N\right|^2$ and $\left|H_2\right|^2\left|N\right|^2$
\begin{eqnarray}
a_{ij}&=&\left.\frac{\partial^2
V_{eff}}{\partial\varepsilon_i\partial\varepsilon_j}\right|_0=
\left.\frac{\partial^2
V_{tree}}{\partial\varepsilon_i\partial\varepsilon_j}\right|_0\nonumber\\
&+& \sum_{a}\left[\frac{\partial V_1}{\partial m_a^4}
\frac{\partial m_a^2}{\partial\varepsilon_i}
\frac{\partial m_a^2}{\partial\varepsilon_j}+
\frac{\partial V_1}{\partial m_a^2}\frac{\partial^2 m_a^2}{
\partial\varepsilon_i\partial\varepsilon_j}\right]_0.\nonumber
\end{eqnarray}
In the above expressions the subscript 0 indicate that all the
derivatives must be evaluated for vanishing fields and the sum over the
index $a$ includes all the particles contributing to $V_1$. The
complete details on how to calculate $\partial m_a^2/\partial
\varepsilon_i$, $\partial m_a^2/\partial
\epsilon_i$,
$(\partial^2 m_a^2/\partial
\varepsilon_i\partial \varepsilon_j)$ and $(\partial^2 m_a^2/\partial
\epsilon_i\partial \epsilon_j)$ are given in the Appendix.

Let's now consider the zero temperature contribution to $V_1$,
$V_1^{T=0}$. In the $\overline{DR}$ scheme of renormalization, it reads
\begin{equation}
V_1^{T=0}=\frac{1}{64\pi^2}{\rm Str}\left\{
{\cal M}^4(\phi)\left[{\rm ln}\frac{{\cal
M}^2(\phi)}{Q^2}-\frac{3}{2}\right]\right\},
\end{equation}
where ${\cal M}^2(\phi)$, with $\phi\equiv \left(H_1^0,H_2^0,N\right)$,
is the field dependent squared mass matrix, the supertrace Str properly
counts the degree of freedom, $Q$ is the renormalization point and the
$Q^2$ dependence in eq. (17) is compensated by that of the renormalized
parameters, so that the full effective potential is independent of $Q^2$
up to the next-to-leading order.

If we consider in $V_1^{T=0}$ only the largest contributions \cite{bb},
namely
those coming from the top and the two stops with masses
\[
m_t^2=h_t^2 \left|H_2^0\right|^2,\:\:\:\: m_{\tilde{t}_L \tilde{t}_R}^2=
M_S^2+h_t^2 \left|H_2^0\right|^2,
\]
where we have taken a common soft mass term $M_S$ for the left and right
stop, one can show that the $B_5$ parameter receives, at the minimum,
a large and {\it positive} correction proportional to $h_t^4\:
{\rm ln}\left(M_S^2/m_t^2\right)$ \cite{bb}. This is the reason why,
when large zero temperature one-loop corrections to $V_{tree}$ coming
from the top-stop sector are considered, one can achieve SCPB only for large
values of $M_S$ \cite{bb}. This result does not contradict the
Georgi-Pais theorem \cite{gp} which is based on the assumption that
one-loop corrections to $V_{tree}$ are small. Indeed, here one-loop
corrections are large enough to turn the CP violating saddle point
into a minimum. Note also that in the limit $\lambda\rightarrow 0$,
the $N$ singlet tends to decouple from the $\left(H_1,H_2\right)$ sector
and a CP violating minimum is found  for
$\left(\varphi_1,\varphi_3\right)=(\pm\pi/4,\pm \pi/2)$ for
$A_{k}<0$ and $A_k>0$, respectively. In such a case, a Peccei-Quinn like
symmetry is present in the superpotential involving $H_1$ and $H_2$ and
it corresponds to the flat direction and then to the massless (at the
tree level)
pseudoscalar boson predicted by the Georgi-Pais theorem. This is the
reason why in ref. \cite{bb} CP violating minima are found only for
small values of $\lambda$.

Let's now come to the renormalization of the $D$
coefficients at finite temperature.

{\large{\bf 3. One-loop Corrections at Finite Temperature}}

\vspace{0.8cm}

The one-loop effective potential at finite temperature is given by
\cite{ft}
\begin{eqnarray}
V_{eff}&=& V_{tree}+V_{1}^{T=0}+V_{1}^{T\neq 0},\nonumber\\
\left(V_{1}^{T=0}+V_{1}^{T\neq 0}\right)_{fer}&=&-
\sum_i n_{i,f}\left[\frac{m_i(\phi)^2T^2}{48}+
\frac{m_i(\phi)^4}{64\pi^2}\left({\rm ln}\frac{Q^2}{A_f
T^2}+\frac{3}{2}\right)\right],\nonumber\\
\left(V_{1}^{T=0}+V_{1}^{T\neq 0}\right)_{bos}&=&
\sum_i n_{i,b}\left[\frac{m_i(\phi)^2T^2}{24}-
\frac{m_i(\phi)^4}{64\pi^2}\left({\rm ln}\frac{Q^2}{A_b
T^2}+\frac{3}{2}\right)\right.\nonumber\\
&-&\left.\frac{T}{12\pi}\left(m_i(\phi)^2+\Pi_i\right)^{3/2}\right],
\end{eqnarray}
where $A_b=16A_f=16\:\pi^2\left(3/2-2\gamma_E\right)$, $\gamma_E$
being the Euler constant, and $n_{i,f(b)}$ counts the effective fermionic
(bosonic) degrees of freedom. Note that in the bosonic part $\Pi_i$
denotes the thermal polarization mass for bosons contributing to the
Debye mas \cite{de}. It arises when one resums at least the leading
infrared-dominated higher-loop contributions to $V_{1}^{T\neq 0}$,
associated to the so called daisy diagrams \cite{mariano} whose
inclusions amounts to a resummation to all orders in $\alpha\sim
(g^2/2\pi)(T^2/m^2)$, where $g$ and $m$ denote the generic gauge coupling
and mass respectively, and to neglect subleading contributions
controlled by the parameter $\beta\sim (g^2/2\pi)(T/m)$.

Note also that in eq. (18) we have made use of the high temperature
expansion for $V_{eff}$ which turns to be a very good approximation
for $m_{if(b)}(\phi)\simless 1.6\: (2.2)\: T$ \cite{hall}.

Accordingly to what explained in the previous Subsection, the
corrections to the coefficients $a_i$ and $a_{ij}$ result as follow

$\bullet$ dimension three operators:
2 Re$\left(N^3\right)$, 2 Re$\left(H_1H_2N^{* 2}\right)$ and
2 Re$\left(H_1H_2N\right)$
\begin{eqnarray}
b_i&=&\left.\frac{\partial V_{eff}}{\partial \epsilon_i}\right|_0=
\left.\frac{\partial V_{tree}}{\partial \epsilon_i}\right|_0
+\sum_{a,fer}n_{a,f}\frac{\partial m_a^2}{\partial\epsilon_i}\left[
\frac{T^2}{48}\right.\nonumber\\
&+&\left.\frac{m_a^2}{32\pi^2}\left({\rm ln}\frac{Q^2}{A_f T^2}+
\frac{3}{2}\right)\right]_0+
\sum_{a,bos}n_{a,b}\frac{\partial m_a^2}{\partial\epsilon_i}
\left[\frac{T^2}{24}\right.\nonumber\\
&-&\left.\frac{m_a^2}{32\pi^2}\left({\rm ln}\frac{Q^2}{A_b T^2}+
\frac{3}{2}\right)-\frac{T}{8\pi}\left(m_a^2+\Pi_a\right)^{1/2}\right]_0,
\end{eqnarray}

$\bullet$ dimension four operators: $\left|H_1\right|^4$,
$\left|H_2\right|^4$, $\left|N\right|^4$,
$\left|H_1\right|^2\left|H_2\right|^2$,
$\left|H_1\right|^2\left|N\right|^2$ and $\left|H_2\right|^2\left|N\right|^2$
\begin{eqnarray}
a_{ij}&=&\left.\frac{\partial^2
V_{eff}}{\partial\varepsilon_i\partial\varepsilon_j}\right|_0=
\left.\frac{\partial^2
V_{tree}}{\partial\varepsilon_i\partial\varepsilon_j}\right|_0\nonumber\\
&+&\sum_{a,fer} n_{a,f}\left\{\frac{1}{32\pi^2}
\frac{\partial m_a^2}{\partial\varepsilon_i}
\frac{\partial m_a^2}{\partial \varepsilon_j}
\left({\rm ln}\frac{Q^2}{A_f T^2}+
\frac{3}{2}\right)\right.\nonumber\\
&+&\left.\frac{\partial^2 m_a^2}{\partial\varepsilon_i\partial\varepsilon_j}
\left[
\frac{T^2}{48}+
\frac{m_a^2}{32\pi^2}\left({\rm ln}\frac{Q^2}{A_f T^2}+
\frac{3}{2}\right)\right]\right\}_0\nonumber\\
&+&\sum_{a,bos}n_{a,b}\left\{\frac{\partial m_a^2}{\partial\varepsilon_i}
\frac{\partial m_a^2}{\partial \varepsilon_j}\left[
-\frac{1}{32\pi^2}\left({\rm ln}\frac{Q^2}{A_b T^2}+
\frac{3}{2}\right)-\frac{T}{16\pi}\left(m_a^2+\Pi_a\right)^{-1/2}\right]
\right.\nonumber\\
&+&\left.\frac{\partial^2 m_a^2}{\partial\varepsilon_i\partial\varepsilon_j}
\left[\frac{T^2}{24}
\frac{m_a^2}{32\pi^2}\left({\rm ln}\frac{Q^2}{A_b T^2}+
\frac{3}{2}\right)-\frac{T}{8\pi}\left(m_a^2+\Pi_a\right)^{1/2}\right]
\right\}_0.
\end{eqnarray}

Working in the 't Hooft-Landau gauge and in the $\overline{DR}$
scheme, in the bosonic part we have summed over gauge bosons, stops,
neutral and charged Higgs scalars, whereas in the fermionic part we have
summed over top, neutralinos and charginos. Again we refer the reader to
the Appendix for the complete analysis of all the technical points.

The coefficients $b_i$ and $a_{ij}$ given in the above eqs. (19) and
(20) are the ones which determine the $D$ coefficients in the eq. (6)
and then determine whether CP may be spontaneously broken at finite
temperature.

In the next Section we shall focus on the results about the SCPB
obtained when finite temperature corrections are taken into account
showing that the picture can be considerably different  from the case in
which only $T=0$ corrections are considered. In particular, we shall
address the question whether it is possible to have SCPB at finite
temperaure in regions of the parameter space which correspond to
CP conserving minima at $T=0$, thus avoiding any limits coming from
having very light Higgs bosons or electric dipole moment of the electron
and neutron close to their experimental bounds.

{\large{\bf 4. Results and Discussion}}

\vspace{0.8cm}

As we have stressed in the Subsection {\bf 2.1}, to have SCPB the
corrections to the tree level potential should act predominantly on the
$B_5$ term, see eq. (15). As a matter of fact, since $B_5<0$ (when
$r<-1/3$) at the tree level, the corrections to $V_{eff}$ should be
large enough to trigger positive values for $B_5$. This is what happens
in the case of $T=0$ top-stop large one-loop corrections, as pointed out
in the Subsection {\bf 2.2}.

In the case of finite temperature corrections to $V_{tree}$, we have
proceeded as follows. First of all, we have fixed the set of the
parameters given by
$(\lambda,k,M_{\tilde{u}},m_N^2,m_1^2,\tan\beta,A_t)$, where
$M_{\tilde{u}}$ is the soft breaking mass for the stop right and $A_t$
is the trilinear soft breaking mass relative to the $h_t\hat{Q}\hat{H_2}
\hat{U}^c$ term in the superpotential. Imposing the minimization
conditions at $T=0$
\begin{equation}
\frac{\partial V_{tree}}{\partial \phi_i}+
\frac{\partial V_{1}^{T=0}}{\partial \phi_i}=0,\:\:\:\:
\left(\phi_1=v_1,\phi_2=v_2,\phi_3=x\right),
\end{equation}
allows one to express the parameters $(A_k,A_{\lambda},m_2^2)$ as
functions of the already fixed parameters and of two free parameters
which we have chosen to be $x$ and $M_{\tilde{q}}$, the latter being the
soft breaking mass of the stop left. In the minimization conditions we
have included only the relevant contributions to $V_{1}^{T=0}$, namely
those coming from the top-stop sector. The corresponding bottom
contributions are negligible for $\tan\beta\simless 20$ and the gauge
sector does not  play any important role. This also holds for the
extended Higgs sector since the Yukawa couplings $\lambda$ and $k$
are taken to be small as suggested by the requirement that they remain
in the perturbative regime up to a large scale (say $10^{16}$ GeV)
\cite{n} or to forbid the breaking of electromagnetism \cite{pok}. Note,
however, that even if the Higgs sector does not play any role in the
$T=0$ corrections, it will be fundamental in the finite temperature
one-loop contributions as we shall explain later.

In Figs. 1a) and 1b) are displayed the plots of the experimentally
allowed regions in the $(M_{\tilde{q}},x)$ plane for fixed values of the
other parameters\footnote{Very recently Ellwanger, R. de Traunbenberg
and Savoy \cite{n} have scanned the complete parameter space searching
for the allowed region compatible with different constraints. In
particular, they have found that only very large values of $x$ are
permitted, $x\simgreat 800$ GeV. On the other hand this result
is based on different assumptions, {\it e.g.}
universality in the soft supersymmetry breaking at the GUT scale. As a
consequence,
we believe that it is safe to consider also smaller values for $x$.}.

The first constraint is that the lightest Higgs CP even particle
$h^0$ has not been produced in the decay $Z^0\rightarrow Z^{0*}+h^0$.
This gives the conservative bound $m_{h^0}\simgreat 60$ GeV \cite{lep},
dashed line (the exact bound depends on the coupling of $h^0$
to $Z^0$, $R_{Z^0 Z^0 h^0}$,  and then it is weaker,
$m_{h^0}\simgreat R_{Z^0 Z^0 h^0}^2\: 60$ GeV). In addition, the lightest
pseudoscalar $A^{0}$ should be heavier than $\sim$ 20 GeV
\cite{lepp} and the dashed-dot line corresponds to $m_{A^0}=40$ GeV.
Note that the allowed regions correspond to $T=0$ CP conserving minima.

When finite temperature corrections are added to $V_{tree}$, one can
define the critical temperature $T_c$ as the value of $T$ at which the
origin of the field space becomes a saddle point for the effective
potential,
\begin{equation}
{\rm Det}\left[{\cal M}_S^{2,T\neq 0}(T_c)\right]_{\phi=0}=0,
\end{equation}
where the effective mass matrix is given by the second derivatives
of the full one-loop finite temperature potential with respect to the
scalar fields. In the origin of the field space and in the basis
$({\rm Re}H_1^0/\sqrt{2},{\rm Re}H_2^0/\sqrt{2},{\rm Re}N/\sqrt{2})$, it
reads
\begin{eqnarray}
{\cal M}_S^{2,T\neq 0}(T_c)&=&{\rm Diag}\left(\overline{m}_1^2,
\overline{m}_2^2,\overline{m}_N^2\right),\nonumber\\
\overline{m}_1^2&=& m_1^2+\frac{1}{8}\left(3g^2+g^{\prime 2}+
\frac{4}{3}\lambda^2\right)T^2,\nonumber\\
\overline{m}_2^2&=&m_2^2+\frac{1}{8}\left(3g^2+g^{\prime 2}
+6 h_t^2+\frac{4}{3}\lambda^2\right)T^2,\nonumber\\
\overline{m}_N^2&=&m_N^2+\frac{1}{3}\left(\lambda^2+k^2\right)T^2.
\end{eqnarray}
The critical temperature is then given by the highest temperature for
which one of the $\overline{m}^2_i$ $(i=1,2,N)$ vanishes. Obviously, the
effective potential becomes flat at the origin only along
directions corresponding to negative soft masses. Due to the heavy top
\cite{cdf} $m_2^2$ can run to negative values at low energy, whereas,
for small $\lambda$ and $k$, both $m_1^2$ and $m_N^2$ remain positive.
As a consequence, the EWPT is expected to occur first along the $H_2^0$
direction and immediately afterwards nonvanishing VEV's for $H_1^0$ and
$N$ are driven by the $A_\lambda$ term \cite{max}.

In Figs. 1a) and 1b) we have fixed the temperature at $T_c=150$ GeV and shown
the curve corresponding to $\overline{m}_2^2(T_c)=0$, solid line. The
points in the $(M_{\tilde{q}},x)$ plane lying on the curve are then the
points for which $T_c=$ 150 GeV. Indeed, since the EWPT is known to
be first order, it occurs when $\overline{m}_2^2$ is still positive,
{\it i.e.} at a temperature $T^*$ higher than $T_c$: since all the points
in the $(M_{\tilde{q}},x)$ plane below the solid line correspond to
$\overline{m}_2^2>0$, then they correspond to the region of the
parameter space where the EWPT occurs at temperatures smaller or equal
to 150 GeV.

We want to point out that the contribution from the Higgs sector, where
plasma effects are crucial to make the effective potential real at the
origin, does have an infrared singularity proportional to
$ (T/\overline{m}_2)$,
responsible for the failure of perturbative expansion \cite{ft} for
values of $(M_{\tilde{q}},x)$ such that $\overline{m}_2^2=0$. The curve
$\overline{m}_2^2=0$ is then representing an upper bound contour, for
fixed $T$, in the $(M_{\tilde{q}},x)$ plane which severly contrains the
region of the parameters. We have also checked that
in the experimentally allowed regions below the curve
$\overline{m}_2^2=0$ and for the choice of $(M_{\tilde{q}},x)$ adopted
for figures 2a) and 2b) (see below),
the theory remais perturbative in the sense that the perturbative
expansion $\beta\sim g^2 (T/2\pi\:\overline{m}_2)$ remains smaller than 1.

Once the temperature has fallen down to the tunneling temperature $T^*$,
the EWPT proceeds via bubble nucleation. In the centre of the bubbles,
to avoid the wiping out of the baryon asymmetry by anomalous
interactions, we have to impose that $\tilde{v}(T^*)=
\sqrt{v_1^2(T^*)+v_2^2(T^*)}\simgreat T^*$, which is easily satisfied in
the model under consideration \cite{max}. Even if the SCPB {\it in the
bubble} is not interesting as far as the generation of the baryon
asymmetry is concerned\footnote{In the usual scenarios \cite{nelson}
for the generation of BAU at the EWPT, the only source of baryon
violations lies in sphaleron interactions whose rate is imposed to
be small enough in the propagating bubble.}, it is however interesting
{\it per se`} asking whether the system can tunnel first to a CP
violating minimum and then, as the temperature decreases,
 reach the CP conserving minimum which represents our initial
condition at $T=0$.
\vskip 0.8 cm

{\bf 4.1. SCPB inside the Bubble}

\vspace{0.8cm}

Scanning several sets of points in the allowed region in the
$(M_{\tilde{q}},x)$ plane, we have numerically minimized the effective
potential and looked for portions in the plane where conditions (15)
could be satisfied. This research has turned out to be fruitless in the
sense that very small and, in practice, negligible, allowed regions in
the $(M_{\tilde{q}},x)$ plane have shown up. Even if this result might
be quite surprising at a first sight, it can be explained as the effect
of different phenomena happening at finite temperature.

We know that
one-loop corrections to $V_{eff}$ should trigger positive values for
$B_5$ in order to have SCPB. First of all, we have seen that at zero
temperature the top-stop sector gives  a large positive
contribution to $B_5$ proportional to $h_t^4\:{\rm ln}(M_S^2/m_t^2)$ for large
$M_S$. At finite temperature, the corresponding contribution
is reduced to $h_t^4\:{\rm ln}(A_b/A_f)$.

Secondly, the Higgs sector plays a crucial role in determining the sign
of $B_5$. Using eqs. (6), (13), the formalism developed
in the Appendix and taking into account that
near the critical temperature $\overline{m}_2\ll\overline{m}_1,
\overline{m}_N$, one can realizes that the corrections to $B_5$
are dominated by negative terms such as $-(T/\overline{m}_2)(\partial
\overline{m}_2/\partial\left|H_{i}^0\right|^2)^2$, with $i=1,2$ and by
$-(T/\overline{m}_2)(\partial
\overline{m}_2/\partial\left|H_{1}^0\right|^2)(\partial
\overline{m}_2/\partial\left|H_{2}^0\right|^2)$, so that the
largest
Higgs contribution to $B_5$, $\Delta B_{5,max}$, is
\begin{equation}
\Delta B_{5,max}=-\frac{1}{8\pi}\frac{T}{\overline{m}_2}\frac{1}{16}
\left(\frac{25}{4}g^4+\frac{30}{4}g^{\prime 4}-\frac{19}{2}
g^2g^{\prime 2}\right)<0.
\end{equation}
Thus, $B_5$ receives a large and ${\it negative}$ contribution from the
one-loop corrections at finite temperature from the Higgs scalar sector.
Since this contribution turns out to be, in absolute values, the largest
among the different particles, this prevents $B_5$ to become
positive and, consequently, the matrix $B$ to be definite positive.
This is the main reason why SCPB cannot occur
in the centre of the propagating bubble.

\vskip 0.8 cm
{\bf 4.2. SCPB inside the Bubble Wall}

\vspace{0.8cm}

Fortunately, the situation changes when one moves away from the centre
of the bubble towards the bubble wall. As a matter of fact, since the
EWPT is of the first order, the temperature keeps constant until the
Universe is in broken phase, whereas $v_1$, $v_2$ and $x$ change their
values from zero to $v_1(T^*)$, $v_2(T^*)$ and $x(T^*)$, respectively,
when a bubble wall passes through a fixed point in space. Incidentally,
we define the bubble wall as the region in which sphalerons are active,
that is in which $0\simless \tilde{v}(T^*)\simless T^*$ and we remind
the reader that, since here the BAU can be created through, {\it e.g.},
the reflection baryogenesis mechanism \cite{nelson}, it is just in this
region that one needs SCPB.

To describe what is happening inside the bubble wall one should  solve a
system of differential equations involving $v_1$, $v_2$, $x$,
$\varphi_1$ and $\varphi_3$ at $T^*$. For instance, the equation for
$\varphi_3$, should read
\begin{equation}
\frac{d}{dz}\left[x^2(z,T^*)\frac{d}{dz}\varphi_3(z,T^*)\right]+
\frac{4}{9}\frac{\partial V_{eff}}{\partial\varphi_3(z,T^*)}=0,
\end{equation}
where we are approximating the bubble wall to an infinite plane propagating
perpendicularly to the $z$ axis with a width $L_w$. Nevertheless, one can
make the analysis much simpler envisaging the following situation.
In the unbroken phase and up to the edge of the advancing bubble wall,
at $z=z_w$,
$v_1$ and $v_2$ are constantly equal to zero and then there is no SCPB;
in the broken phase and up to the other edge of the bubble wall,
at $z=z_w+L_w$,
$v_1$ and $v_2$ are nearly constantly equal to their minimum values at the
centre of the bubble and, as shown before, again no SCPB is occuring;
finally from inside to outside of the bubble wall, the VEV's are
decreasing from their values inside the bubble to zero.

The key point now is that {\it in} the bubble wall, where equations like
eq. (25) are valid, the VEV's $v_1(z,T^*)$, $v_2(z,T^*)$ and
$x(z,T^*)$ do not have to satisfy any longer the severe constraint
$B_5>0$ coming from the requirment of a mass matrix with positive determinant,
but one can
simply impose that at the two edges of the bubble wall,
where, {\it e.g.}, eq. (25) reduces to $\partial V_{eff}/\partial
\varphi_3(z,T^*)=0$, $\cos 2\varphi_3(z_w+L_w,T^*)>1$ and
$\cos 2\varphi_3(z_w,T^*)<-1$. So doing, even if the solution of eq.
(25) is not known, one is assured that maximal SCPB is occuring inside
the bubble wall, that is an observer in a fixed point in space
would experience a change $\Delta\varphi_3=\pi/2$.
Imposing $\cos 2\varphi_3(z_w+L_w,T^*)>1$ and
$\cos 2 \varphi_3(z_w,T^*)<-1$ at the edges of the bubble wall is
equivalent to impose
\begin{equation}
0< \tilde{v}_{-}<\tilde{v}(T^*)<\tilde{v}_{+}<\tilde{v}(z_w+L_w,T^*),
\end{equation}
where $\tilde{v}_{\pm}$ are found imposing $\cos 2\varphi_3(T^*)=\pm 1$,
respectively, and are given by
\begin{equation}
\tilde{v}_{\pm}=\frac{\sqrt{2}}{\left|\sin 2\beta\right|^{1/2}}
\left[\pm\frac{2 \:D_{12}^2}{D_7 D_{11}}\: x^3(z,T^*)
+\frac{D_{12}^2}{D_{7}^2}\:x^2(z,T^*)+\frac{D_{12}^2}{D_{11}^2}\:x^4(z,T^*)
\right]^{1/4}.
\end{equation}
At $\tilde{v}(z,T^*)=\tilde{v}_{\pm}$ eq. (25) is then equivalent
to $\partial V_{eff}/\partial
\varphi_3(z,T^*)=0$ and also eq. (10) must be satisfied.
Naturally, similar relations are valid for
$\tilde{v}^2(z,T^*)$ and $x(z,T^*)$ once one
imposes similar conditions on the angles
$2\varphi_1$ and $2(\varphi_1-\varphi_3)$.

In Figs. 2a) (2b)) we present the region in the $(\tilde{v}(z,T^*),
x(z,T^*))$ plane corresponding to the conditions (26-27)
for the choice
$x=$ 600 (800) GeV, $m_N=10$, 300 GeV and $M_{\tilde{q}}=200\: (300)
$ GeV. The points
inside the regions I and II correspond to values of $\tilde{v}(z,T^*)$
and $x(z,T^*)$ for which a maximal SCPB is occuring in the bubble wall,
{\it i.e}, at least one of the angles $2\varphi_3$, $2\varphi_1$ and
$2(\varphi_1-\varphi_3)$
change by an amount $\pi$ inside the bubble wall.

As pointed out in ref. \cite{cpr}, the only presence of SCPB inside the
bubble wall is not sufficient to create a net BAU. Indeed, if CP is
spontaneously broken, any phase, let's call it $\delta$, can take two
opposite values, corresponding to two exactly degenerate vacua, since
the effective potential depends only on $\cos\delta$. This degeneracy
between the two vacua $\pm\delta$ would induce an equal number of
nucleated bubbles carrying phases with opposite signs, which in turn
generate baryon asymmetries of opposite signs and an overall BAU equal
to zero when averaging over the entire volume of the Universe. However,
the presence of a very small and explicit phase in the effective
potential, for instance in $r$, of order of $10^{-6}-10^{-5}$, is
sufficient to lift the degeneracy leading to a difference between
nucleation rates of the two kind of bubbles and to a baryon asymmetry of
the right order of magnitude \cite{cpr}. This is due to the fact that,
being the nucleation rate proportional to ${\rm exp}(-\Delta F/T)$,
where $\Delta F$ is the difference in free energy between the two vacua
with phases $\pm\delta$, it is very sensible to even small changes in
$\Delta F$. The smallness of the explicit phase does not change
quantitatively our conclusions on SCPB and give rise to negligible
contributions to the neutron electric dipole moment.
% Finally, we note
%that, even if small explicit CP violating phases are present in the
%potential, the $Z_3$ symmetry is still present unless nonrenormalizable
%operators like the one mentioned previously, $N^4/M$, are introduced
%in the superpotential. This residual symmetry generates three
%equivalent vacua, giving rise to  This fact again would give rise to a
%%vanishing
%BAU since, when going from one of the three  values of $\theta_3$ for which
%$\cos\theta_3=+1$ to one of the three for which $\cos\theta_3=-1$, one can
%%have
%the same absolute values of $\Delta\theta_3$, but opposite in sign. From
%what discussed in
%ref. \cite{cpr} one can argue that a net BAU can be generated if the
%mass $M$ of the operator $N^4/M$ is around the GUT scale.
\vskip 0.8 cm
{\large{\bf 5. Conclusions}}
\vspace{0.8cm}\\

In the present paper we have investigated the possibility of SCPB at
finite temperature in the MSSM. After having performed a systematical
analysis of the renormalization of the operators present in the
potential, we have shown that SCPB at finite temperature cannot occur
inside the propagating bubble walls which appear after the EWPT. This is
due to the fact that the Higgs contribution turns out to be crucial and
pushing towards the wrong direction.

SCPB can occur inside the bubble wall, which is the interesting region
as far as the generaton of the BAU is concerned, without any fine-tuning
and in regions of the parameter space which corespond at $T=0$ to no
SCPB, {\it i.e.} to regions where no constraints are present coming from
experimental bounds on the electron and/or neutron electric dipole
moment and from searching for the lightest Higgs bosons at LEP.

Very small explicit phases are necessary to generate a nonvanishing
BAU and in any case they give rise to an amount of
CP violation which is well below
the current experimental limits.
\vskip 0.3 cm
{\bf 2.1. Acknowledgements}
\vspace{0.3cm}\\
It is a pleasure to than R. Hempfling for useful and enlightening
discussions.
\vspace{0.3cm}
\setcounter{secnumdepth}{0}
\section{Appendix.}
%\centerline{\Large\bf Appendix.}
\setcounter{section}{1}
\renewcommand{\thesection}{\Alph{section}}
\renewcommand{\theequation}{\thesection . \arabic{equation}}
\vspace{.5 truecm}
\setcounter{equation}{0}

In this Appendix we want to give a complete description of all the
techniques necessary to calculate the corrections to the coefficients
$a_i$ and $a_{ij}$. Note that this description is quite general and does
not depend at all on the particular model we are working with.

As explained in the text, to calculate the renormalized
$a_i$ and $a_{ij}$ coefficients, we need to calculate $\partial
m_a^2/\partial\varepsilon_i$ and $(\partial^2 m_a^2/\partial\varepsilon_i
\partial\varepsilon_j)$, $m_a^2$ being the field dependent mass
eigenvalues of the particles contributing to the effective potential.
Since $m_a^2$ are by definition the squared mass eigenvalues of an
$N$x$N$ hermitian mass matrix ${\cal M}^2_N$, they obey the following
equation
\begin{equation}
{\cal U}\left[m_a^2(\phi),c_n(\phi)\right]\equiv{\rm Det}
\left[{\cal M}^2_N(\phi)-m_a^2(\phi)\right]=\sum_{n=0}^{N}
c_{n}(\phi)m_{a}^{2n}(\phi)=0.
\end{equation}
Let's analyze first the case for a {\it non degenerate} mass matrix
${\cal M}_N^2$. Taking the derivative of both sides of eq. (A.1) with
respect to the gauge invariant and independent operator $\varepsilon_i$,
one can easily finds that (analogous expressions are valid for the
dimension three operators $\epsilon_i$)
\begin{equation}
\frac{\partial m_a^2}{\partial\varepsilon_i}=-\frac{
\sum_{n=0}^{N}m_{a}^{2n}\frac{\partial c_n}{\partial\varepsilon_i}}{
\sum_{n=0}^{N}n c_n m_{a}^{2(n-1)}},
\end{equation}
while, taking the second derivatives with respect to the operators
$\varepsilon_i$ and $\varepsilon_j$, one finds
\begin{eqnarray}
\frac{\partial^2 m_a^2}{\partial\varepsilon_i\partial\varepsilon_j}
&=&-\frac{1}{
\sum_{n=0}^{N}n c_n m_{a}^{2(n-1)}}
\sum_{n=0}^{N}\left[n(n-1)c_n\frac{\partial m_a^2}{\partial\varepsilon_i}
\frac{\partial m_a^2}{\partial\varepsilon_j}m_a^{2(n-2)}\right.\nonumber\\
&+&\left.m_a^{2n}\frac{\partial^2
c_n}{\partial\varepsilon_i\partial\varepsilon_j}+
nm_{a}^{2(n-1)}\left(\frac{\partial c_n}{\partial\varepsilon_i}
\frac{\partial m_a^2}{\partial\varepsilon_j}+
\frac{\partial c_n}{\partial\varepsilon_j}
\frac{\partial m_a^2}{\partial\varepsilon_i}\right)\right].
\end{eqnarray}
In the case of {\it degenerate} squared mass matrices, {\it e.g.}
the 6x6 squared mass matrix for the neutral Higgs bosons (see below),
one cannot use the equation
\begin{equation}
\frac{\partial{\cal U}}{\partial\varepsilon_i}=
\frac{\partial{\cal U}}{\partial m_a^2}
\frac{\partial m_a^2}{\partial\varepsilon_i}+
\sum_{n=0}^N m_a^{2n}\frac{\partial c_n}{\partial\varepsilon_i}=0
\end{equation}
to invert in favour of $(\partial m_a^2/\partial\varepsilon_i)$ since
now $(\partial{\cal U}/\partial m_a^2)$ is vanishing. Nevertheless, for
$m_a^2$ eigenvalues with degree of degeneracy equal to two, one can take
the derivative of eq. (A.4) with respect to $m_a^2$ and $\partial^2
m_a^2/\partial
\varepsilon_i\varepsilon_j$then obtains
\begin{equation}
\frac{\partial m_a^2}{\partial\varepsilon_i}=-\frac{
\sum_{n=0}^{N}n m_{a}^{2(n-1)}\frac{\partial c_n}{\partial \varepsilon_i}}{
\sum_{n=0}^{N}n(n-1) c_n m_{a}^{2(n-2)}}
\end{equation}
and
\begin{eqnarray}
\frac{\partial^2 m_a^2}{\partial\varepsilon_i\partial\varepsilon_j}
&=&-\frac{1}{
\sum_{n=0}^{N}n(n-1) c_n m_{a}^{2(n-2)}}
\sum_{n=0}^{N}\left[n(n-1)(n-2)c_n\frac{\partial m_a^{2}}
{\partial\varepsilon_i}
\frac{\partial m_a^2}{\partial\varepsilon_j}m_a^{2(n-3)}\right.\nonumber\\
&+&\left.n m_a^{2(n-1)}\frac{\partial^2
c_n}{\partial\varepsilon_i\partial\varepsilon_j}+
n(n-1)m_{a}^{2(n-2)}\left(\frac{\partial c_n}{\partial\varepsilon_i}
\frac{\partial m_a^2}{\partial\varepsilon_j}+
\frac{\partial c_n}{\partial\varepsilon_j}
\frac{\partial m_a^2}{\partial\varepsilon_i}\right)\right].
\end{eqnarray}
What one really needs to calculate the corrections to the coefficients
$a_i$ and $a_{ij}$, which in their turn determine the possibility of
having SCPB at finite temperature, are then the coefficients
for each squared mass matrix whose eigenstates contribute to the
effective potential. In the following we give all the details about the
properties of the particles which must be included in $V_{eff}$ and from
which the coefficients $c_n(\phi)$ can be extracted.

$\bullet$ Top: $n_t=-12$, $m_t^2=h_t^2\left|H_2^0\right|^2$.

$\bullet$ Stop: $n_{\tilde{t}_1}=n_{\tilde{t}_1}=6$; the squared
mass matrix is
\begin{equation}
{\cal M}^2_{\tilde{t}}=
\left(\begin{array}{cc}
m_{LL}^2& m_{LR}^2\\
m_{LR}^{*2}&m_{RR}^2
\end{array}\right),
\end{equation}
where
\begin{eqnarray}
m_{LL}^2&=& M_{\tilde{q}}^2+\Pi_{\tilde{q}}+h_t^2\left|H_2^0\right|^2
+\left(\frac{g^2}{12}-\frac{g^{\prime 2}}{4}\right)\left(
\left|H_2^0\right|^2-\left|H_1^0\right|^2\right),\nonumber\\
m_{RR}^2&=& M_{\tilde{u}}^2+\Pi_{\tilde{u}}+h_t^2\left|H_2^0\right|^2
-\frac{g^{\prime 2}}{3}\left(
\left|H_2^0\right|^2-\left|H_1^0\right|^2\right),\nonumber\\
M_{LR}^2&=&h_t\left(A_t H_2^0+\lambda N^* H_1^{*0}\right),
\end{eqnarray}
and \cite{rec}
\begin{equation}
\Pi_{\tilde{q}}=\Pi_{\tilde{u}}\simeq\frac{4}{9}g_s^2T^2
\end{equation}
are the thermal polarization squared masses for the squarks
calculated in the limit
of heavy gluinos.

$\bullet$ Gauge bosons: $n_{W^{\pm}}=6$, $n_{W^{\pm}_{L}}=2$,
$n_{Z^0}=3$, $n_{Z^0_{L}}=n_{\gamma_{L}}=1$,
\begin{eqnarray}
m_{W^{\pm}}^2&=&\frac{g^2}{2}\left(\left|H_1^0\right|^2
+\left|H_2^0\right|^2\right),\nonumber\\
m_{Z^0}^2&=&\frac{g^2+g^{\prime 2}}{2}\left(\left|H_1^0\right|^2
+\left|H_2^0\right|^2\right),\\
\overline{m}^2_{W^{\pm}_{L}}&=&m_{W^{\pm}}^2+\Pi_{W^{\pm}_{L}},\nonumber\\
\overline{m}^2_{Z^0_{L},\gamma_L}&=&
\frac{1}{2}\left[m_{Z^0}^2+\Pi_{W^{\pm}_{L}}+
\Pi_{B_{L}}\right.\nonumber\\
&\pm&\left.\sqrt{\frac{1}{4}\left[
\frac{g^2-g^{\prime 2}}{2}\left(\left|H_1^0\right|^2
+\left|H_2^0\right|^2\right)+\Pi_{W^{\pm}_{L}}-
\Pi_{B_{L}}\right]^2+\left[\frac{g g^{\prime}}{2}
\left(\left|H_1^0\right|^2
+\left|H_2^0\right|^2\right)\right]^2}\right],\nonumber
\end{eqnarray}
where \cite{rec}
\begin{equation}
\Pi_{W^{\pm}_{L}}=\frac{5}{2}g^2T^2,\:\:\:\:,
\Pi_{B_{L}}=\frac{47}{10}g^{\prime}T^2.
\end{equation}
Note that only the longitudinal components of the gauge bosons do have a
Debye mass proportional to $T^2$.

$\bullet$ Charginos: $n_{\tilde{\chi}^{\pm}}=2$; in the basis
$(-i\lambda^{+},\psi_{H_2}^{1},-i\lambda^{-},\psi_{H_1}^2)$ the mass
matrix is
\begin{equation}
{\cal M}_{\tilde{\chi}^{\pm}}=
\left(\begin{array}{cccc}
0& 0 & M_2 & g H_1^{*0}\\
0& 0 & g H_2^{*0}& \lambda N\\
M_2 & g H_2^{*0}&0 &0\\
g H_1^{*0} & \lambda N & 0 & 0
\end{array}\right),
\end{equation}
where $M_2$ is the gaugino soft mass for the gauge group $SU(2)_L$.

$\bullet$ Neutralinos: $n_{\tilde{\chi}^{0}}=2$; in the basis
$(\tilde{W}^3,\tilde{B}^0,\tilde{H}_1^0,\tilde{H}_2^0,\tilde{N})$
the mass matrix is
\begin{equation}
{\cal M}_{\tilde{\chi}^{0}}=
\left(\begin{array}{ccccc}
M_2 & 0 -\frac{g}{2}H_1^{*0} & \frac{g}{2}H_2^{*0} & 0\\
0 & M_1 & \frac{g^{\prime}}{2}H_1^{*0} &-\frac{g^{\prime}}{2}H_2^{*0}&
0\\
-\frac{g}{2}H_1^{*0}&\frac{g^{\prime}}{2}H_1^{*0}& 0&\lambda N& \lambda
H_2^0\\
\frac{g}{2}H_2^{*0}&-\frac{g^{\prime}}{2}H_2^{*0}&\lambda N &0 & \lambda
H_1^0\\
0 & 0 & \lambda H_2^0 & \lambda H_1^0 & -2k N
\end{array}\right),
\end{equation}
where $M_1$ is the gaugino soft mass for the gauge group $U(1)_Y$.

$\bullet$ Charged Higgs scalars: $n_{H^{\pm}}=2$; in the basis
$(H_1^{-}, H_2^{+*})$ the squared mass matrix reads
\begin{eqnarray}
{\cal M}^2_{H^{\pm},11}&=&\overline{m}_1^2+\lambda^2\left|N\right|^2+
\frac{1}{4}\left(g^2+g^{\prime 2}\right)\left(\left|H_1^0\right|^2
+\left|H_2^0\right|^2\right),\nonumber\\
{\cal M}^2_{H^{\pm},22}&=&\overline{m}_2^2+\lambda^2\left|N\right|^2-
\frac{1}{4}\left(g^2+g^{\prime 2}\right)\left(\left|H_1^0\right|^2
+\left|H_2^0\right|^2\right)+\frac{g^2}{2}\left|
H_1^0\right|^2,\nonumber\\
{\cal M}^2_{H^{\pm},12}&=&{\cal M}^{2*}_{H^{\pm},21}=
\lambda A_{\lambda}N+\lambda k
N^{*2}+\left(\frac{1}{2}g^2-\lambda^2\right)H_{1}^{*0}
H_{2}^{*0}.
\end{eqnarray}

$\bullet$ Neutral Higgs scalars: $n_{H^{0}}=1$; in the basis
$(H_1^0,H_1^{*0},H_2^0,H_2^{*0},N,N^*)$ the squared mass matrix is given
by
\begin{eqnarray}
{\cal M}^2_{H^{0},11}&=&\overline{m}_1^2+\lambda^2\left|N\right|^2+
\lambda^2\left|H_2^0\right|^2+
\frac{1}{2}\left(g^2+g^{\prime 2}\right)\left|H_1^0\right|^2,
-\frac{1}{4}\left(g^2+g^{\prime 2}\right)\left|H_2^0\right|^2,\nonumber\\
{\cal M}^2_{H^{0},12}&=&{\cal M}^{2*}_{H^{0},21}=
{\cal M}^{2}_{H^{0},34}=
\frac{1}{4}\left(g^2+g^{\prime 2}\right)H_1^0H_2^0,\nonumber\\
{\cal M}^2_{H^{0},13}&=&{\cal M}^{2*}_{H^{0},31}=
{\cal M}^{2*}_{H^{0},24}=
\left[\lambda^2-\frac{1}{4}\left(g^2+g^{\prime 2}\right)\right]
H_{1}^{0}H_2^{*0},\nonumber\\
{\cal M}^2_{H^{0},14}&=&{\cal M}^{2*}_{H^{0},41}=
{\cal M}^{2*}_{H^{0},23}=
\left[\lambda^2-\frac{1}{4}\left(g^2+g^{\prime 2}\right)\right]
H_{1}^{0}H_2^{0}-\lambda k N^2-\lambda A_{\lambda}N^*,\nonumber\\
{\cal M}^2_{H^{0},15}&=&{\cal M}^{2*}_{H^{0},51}=
{\cal M}^{2*}_{H^{0},26}=
\lambda^2H_1^0N^*-\lambda k H_{2}^{*0}N,\nonumber\\
{\cal M}^2_{H^{0},16}&=&{\cal M}^{2*}_{H^{0},61}=
{\cal M}^{2*}_{H^{0},25}=
\lambda^2 N H_{1}^0-\lambda A_{\lambda}H_{2}^{*0},\nonumber\\
{\cal M}^2_{H^{0},22}&=&\overline{m}_1^2+\lambda^2\left|N\right|^2+
\lambda^2\left|H_2^0\right|^2+
\frac{1}{2}\left(g^2+g^{\prime 2}\right)\left|H_1^0\right|^2,
-\frac{1}{4}\left(g^2+g^{\prime 2}\right)\left|H_2^0\right|^2,\nonumber\\
{\cal M}^2_{H^{0},33}&=&\overline{m}_2^2+\lambda^2\left|N\right|^2+
\lambda^2\left|H_1^0\right|^2+
\frac{1}{2}\left(g^2+g^{\prime 2}\right)\left|H_2^0\right|^2-
\frac{1}{4}\left(g^2+g^{\prime 2}\right)\left|H_1^0\right|^2,\nonumber\\
{\cal M}^2_{H^{0},35}&=&{\cal M}^{2*}_{H^{0},53}=
{\cal M}^{2*}_{H^{0},46}=
\lambda^2H_2^0 N^*-\lambda k H_{1}^{*0}N,\nonumber\\
{\cal M}^2_{H^{0},36}&=&{\cal M}^{2*}_{H^{0},63}=
{\cal M}^{2*}_{H^{0},45}=
\lambda^2 H_{2}^0-\lambda A_{\lambda}H_{1}^{*0},\nonumber\\
{\cal M}^2_{H^{0},44}&=&\overline{m}_2^2+\lambda^2\left|N\right|^2+
\lambda^2\left|H_1^0\right|^2+
\frac{1}{2}\left(g^2+g^{\prime 2}\right)\left|H_2^0\right|^2-
\frac{1}{4}\left(g^2+g^{\prime 2}\right)\left|H_1^0\right|^2,\nonumber\\
{\cal M}^2_{H^{0},55}&=&\overline{m}_N^2+\lambda^2\left|H_1^0\right|^2+
\lambda^2\left|H_2^0\right|^2+
2k^2\left|N\right|^2,\nonumber\\
{\cal M}^2_{H^{0},56}&=&{\cal M}^{2*}_{H^{0},65}=
2k^2N^{2}+2\lambda k H_{1}^0H_{2}^0-3k A_k N^*,\nonumber\\
{\cal M}^2_{H^{0},66}&=&\overline{m}_N^2+\lambda^2\left|H_1^0\right|^2+
\lambda^2\left|H_2^0\right|^2+
4k^2\left|N\right|^2.
\end{eqnarray}

\newpage
%
%%%%%%%%%%%%%%%%%%--- References
%%%%%%%%%%%%%%%%%%%%%%%%%%%%%%%%%%%%%%%%%%%%%%%%%%%%%%%
\def\MPL #1 #2 #3 {Mod.~Phys.~Lett.~{\bf#1}\ (#3) #2}
\def\NPB #1 #2 #3 {Nucl.~Phys.~{\bf#1}\ (#3) #2}
\def\PLB #1 #2 #3 {Phys.~Lett.~{\bf#1}\ (#3) #2}
\def\PR #1 #2 #3 {Phys.~Rep.~{\bf#1}\ (#3) #2}
\def\PRD #1 #2 #3 {Phys.~Rev.~{\bf#1}\ (#3) #2}
\def\PRL #1 #2 #3 {Phys.~Rev.~Lett.~{\bf#1}\ (#3) #2}
\def\RMP #1 #2 #3 {Rev.~Mod.~Phys.~{\bf#1}\ (#3) #2}
\def\ZP #1 #2 #3 {Z.~Phys.~{\bf#1}\ (#3) #2}

\newpage
%%%%%%%%%%%%%%%%%%%%%%%--- figures
\noindent{\bf Figure Caption}
\begin{itemize}
\item[{\bf Fig. 1a)}]{ For the following values of the parameters:
$\lambda=0.2$, $k=0.4$, $h_t=1$, $M_{\tilde{u}}=M_{\tilde{q}}+10$ GeV,
$m_{N}=10,300$
GeV, $m_1=$ 250 GeV, $\tan\beta=2.5$, $A_t=10$ GeV, $M_1=M_2=150$ GeV
 and $T=150$ GeV, and in the
plane $(x,M_{\tilde{q}})$, the solid
line corresponds to the curve $\overline{m}_2(T)=0$, the dot-dashed line
corresponds to $m_{A^0}=40$ GeV and the dashed line to the conservative
bound $m_{h^0}\simgreat$ 60 GeV. Regions I and II correspond to
$m_{N}=10,300$ GeV, respectively.}
\end{itemize}
\begin{itemize}
\item[{\bf Fig 1b)}]{The same as in Fig. 1a) but for $\lambda=0.3$, $k=$
0.5, $\tan\beta=1.2$,
$m_1=150$ GeV,
$M_1=M_2=250$ GeV.}
\end{itemize}
\begin{itemize}
\item[{\bf Fig. 2a)}] {Values in the plane $(\tilde{v}(z,T^*), x(z,T^*))$
for which a maximal SCPB occurs. Region I and II correspond to the
choice of the parameters given in Fig. 1a) with $x=600$ GeV and
$M_{\tilde{q}}=200$ GeV.}
\end{itemize}
\begin{itemize}
\item[{\bf Fig. 2b)}] {Values in the plane $(\tilde{v}(z,T^*), x(z,T^*))$
for which a maximal SCPB occurs. Region I and II correspond to the
choice of the parameters given in Fig. 1b) with $x=800$ GeV and
$M_{\tilde{q}}=300$ GeV.}
\end{itemize}

\end{document}